\documentclass{aa}        
\usepackage{txfonts}
\usepackage{graphicx,longtable,lscape,natbib,psfig,textcomp,fontenc,amssymb}  
\bibpunct{(}{)}{;}{a}{}{,} 
\newcommand{\ltsima} {$\; \buildrel < \over \sim \;$}  
\newcommand{\gtsima} {$\; \buildrel > \over \sim \;$}  
\newcommand{\lta} {\lower.5ex\hbox{\ltsima}}  
\newcommand{\gta} {\lower.5ex\hbox{\gtsima}}

\begin{document}
\title{Molecular gas and nuclear activity in early-type galaxies: \\any
  link with radio-loudness?\thanks{Based on observations carried out with
  the IRAM 30m telescope. IRAM is supported by INSU/CNRS (France), MPG
  (Germany), and IGN (Spain).}}
  
\titlerunning{Molecular gas and nuclear activity in early-type galaxies}

\authorrunning{R.~D. Baldi et al.}  

\author{Ranieri D. Baldi \inst{1,2,3} 
\and  Marcello Giroletti \inst{4}
\and  Alessandro Capetti \inst{5}
\and  Gabriele Giovannini \inst{4,6}
\and  Viviana Casasola \inst{4}
\and \\ Miguel A. P\'erez-Torres \inst{7}
\and  Nario Kuno \inst{8}}

\institute{
SISSA-ISAS, via Bonomea 265, I-34136 Trieste, Italy 
\and 
Physics Department, The Technion, Haifa 32000, Israel 
\and 
Physics Department, Faculty of Natural Sciences, University of Haifa, Israel
\and
INAF-Istituto di Radio Astronomia, via P. Gobetti 101, I-40129 Bologna, Italy
\and
INAF - Osservatorio Astrofisico di Torino, Via Osservatorio 20,
  I-10025 Pino Torinese, Italy
\and
Dipartimento di Fisica e Astronomia, Universit\`a di Bologna, via Ranzani 1, 40127 Bologna, Italy 
\and
Instituto de Astrof\'{\i}sica de Andaluc\'{\i}a - CSIC, PO Box 3004, 18008 Granada, Spain
\and
Division of Physics, Faculty of Pure and Applied Sciences, University of Tsukuba,  Japan}

\offprints{baldi@ph.technion.ac.il} 

\date{}

  \abstract{
   {}
  
   {\it Aims}. {We want to study the amount of molecular
   gas in a sample of nearby early-type galaxies (ETGs) which host
   low-luminosity Active Galactic Nuclei (AGN). We look for possible differences
   between the radio-loud (RL) and radio-quiet (RQ) AGN.}\\
   {\it Methods}. {We observed the CO(1-0) and CO(2-1)
   spectral lines with the IRAM 30m and NRO 45m telescopes for eight
   galaxies. They belong to a large sample of 37 local ETGs which host
   both RQ and RL AGN. We gather data from the literature for the
   entire sample.}\\
     {\it Results}. {We report the new
   detection of CO(1-0) emission in four galaxies (UGC\,0968,
   UGC\,5617, UGC\,6946, and UGC\,8355) and CO(2-1) emission in two of
   them (UGC\,0968 and UGC\,5617). The CO(2-1)/CO(1-0) ratio in these sources is
   $\sim0.7\pm0.2$. Considering both the new observations and the
   literature, the detection rate of CO in our sample is 55 $\pm$
     9\%, with no statistically significant difference between the
   hosts of RL and RQ AGNs. For all the detected galaxies we converted
   the CO luminosities into the molecular masses, $M_{H_2}$, that
   range from 10$^{6.5}$ to 10$^{8.5}$ M$_{\odot}$, without any
   statistically significant differences between RL and RQ
   galaxies. This suggests that the amount of molecular gas does not
   likely set the radio-loudness of the AGN. Furthermore, despite
     the low statistical significance, the presence of a weak trend
     between the H$_{2}$ mass with various tracers of nuclear activity
     (mainly [O~III] emission line nuclear power) cannot be
     excluded.}
   {}

}

\maketitle

\section{Introduction}


CO is the most abundant molecule in galaxies after H$_{2}$ (CO/H$_{2}$
$\sim$ $6\times10^{-5}$); it is easy to excite, and it radiates
efficiently at frequencies which can be observed fairly easily from
the ground. Rotational transitions of CO are therefore the best
tracers of cold molecular gas in galaxies.


In general, early-type galaxies (ETGs) contain an interstellar medium
(ISM) with components similar to those found in spiral galaxies as
shown by radio, optical, and X-ray observations. However, the detailed
analysis of the gas components of large samples of galaxies has shown
the existence of differences in ISM between galaxies of different
morphological types. In addition, the spatial distribution and the
relative gas fraction of the different phases differ between early and
late-type galaxies. The molecular gas in ETGs seems to be more
centrally concentrated than in later types (e.g.,
\citealt{boker03,komugi08,davis13}). In elliptical galaxies a hot X-ray
radiating coronal halo is the most massive gaseous component with
masses between 10$^8$ and 10$^{10}$ M$_{\odot}$ (e.g.,
\citealt{fabbiano89}). Warm ionized gas with estimated masses of
$10^{3}-10^{5}$ M$_{\odot}$ is found in $\sim$60\% of the observed
ellipticals (e.g.,
\citealt{phillips86,sadler87,burstein97,beuing99,bettoni03}). The
presence of a cold gas component in ellipticals has been initially
inferred from observations of HI and mid- and far-infrared emission
\citep{knapp85,knapp89}. A growing number of ellipticals are found to
contain a molecular gas component with masses ranging from 10$^{7}$ to
10$^{10}$ M$_{\odot}$ (e.g.,
\citealt{lees91,wiklind95,young02,young05,bettoni03,combes07,sage07,young11,osullivan14}).

Moreover, since an ETG can host both RL and RQ AGN, this type of
galaxies appears to be a critical class of objects where a large
variety of nuclear and host properties might help us to understand
their connection. The relative fraction of hot and cold gas phase,
traced by the X-ray emitting coronal emission and the CO molecule present in
the ISM of ETGs, might play an important role in their accretion mode
and their radio-loudness. Recently, \citet{werner14} studied the
properties of the multi-phase gas in ETGs and their role in fueling
radio-mode AGN feedback.

In order to quantify the molecular gas content in the ETGs, we will
take advantage on the substantial knowledge about the multiband
nuclear and host properties of a sample of ETGs described
below. \citet{capetti05} studied the optical surface brightness
profiles based on the HST images of 37 nearby ETGs, by separating the
ETGs on the basis of the inner logarithmic slope $\gamma$ into Core
($\gamma\leq$0.3) and Power-law ($\gamma>$0.5) galaxies (hereafter
CoreG and PlawG, respectively). This sample includes both RL and RQ
AGN. On one hand, the CoreG show flat optical core and host RL AGN,
with multiwavelength nuclear and host properties similar to those of
the FR~Is \citep{fanaroff74}.  On the other hand, PlawG are the
counterparts of the RQ Seyfert galaxies in ETGs at smaller nuclear
luminosities with similar sharp inner optical profiles
\citep{capetti07b}.

In this work we will explore the link between the molecular gas
present in these complete samples of ETGs and the radio-loudness and
the nuclear properties. The structure of this paper is as follows. In
Sect. 2, we describe our observations and the literature data we used
to complete the sample. In Sect. 3, we present the results giving
H$_{2}$ masses of galaxies and the computed CO(2-1)/CO(1-0) ratio.  In
Sect 4. we search for empirical relations between the molecular gas
and the nuclear activity, considering the entire sample of CoreG and
PlawG. In Sect. 5, we summarize the results and draw the conclusions.

We adopt a Hubble constant of H$_{0}$ = 71 km s$^{-1}$ Mpc$^{-1}$ and q$_{0}$
= 0.5. Assuming a different cosmology would not affect our results since the
redshift of the sources considered in this paper is limited to $z=0.01$.

\section{Sample, observations and data}
\label{sample}

The galaxies studied in this work belong to a sample of luminous (B
$<$ 14), nearby (V$_{rec}$ $<$ 3000 km s$^{-1}$) ETGs.
\citet{wrobel91b} performed a VLA survey with a flux cutoff of $\sim$1
mJy at 5 GHz on this sample. For the radio-detected objects,
\citet{capetti05} used archival HST observations, available for 48
objects, to study their surface brightness profiles and to separate
these ETGs following the Nukers scheme \citep{faber97} rather than the
traditional morphological classification (i.e. into E and S0
galaxies). They divided the sample in CoreG and PlawG, characterized
respectively by a flat and a sharp inner logarithmic slope of their
optical brightness profiles.  For 37 galaxies the surface brightness
profile is sufficiently well-behaved to yield a successful
classification. These sources constitute the sample studied in this
work.

Their multiwavelength (radio, optical, X-ray, and emission line)
  properties have been analyzed in a series of papers by
  \citet{capetti05}, \citet{balmaverde06b}, \citet{balmaverde06a},
  \citet{capetti06}, \citet{capetti07b}, \citet{baldi09}, and Baldi \&
  Capetti (in prep.). These studies show that the two optical classes
  of objects correspond to different AGN properties. On one hand, the
  CoreG represent 'miniature radio galaxies', i.e. a scaled-down
  version of FR~Is \citep{baldi09}. They invariably host RL nuclei,
  spectroscopically classified as low-excitation galaxies, with an
  average radio-loudness parameter of R=L$_{\rm 5 \,\ GHz}$/L$_{B}
  \sim$4000, similarly to FR~Is. In addition, they share with the FR~Is
  a common non-thermal AGN jet origin of the radio, X-ray, and optical
  nuclear emission (e.g., \citealt{chiaberge99,balmaverde06b}). On the
  other hand, the PlawG are low-luminosity RQ AGN (on average 100 time
  less RL than the CoreG), mostly with LINER spectra, but in a few
  cases they are also associated with Seyfert optical spectra (Baldi
  \& Capetti in prep.). They show a deficit in radio luminosity at a
  given optical and X-ray luminosity with respect to the CoreG
  \citep{capetti06,capetti07b,kharb12}.  Although the sample is
  detected in the radio band as well in other wavelengths thanks to
  deep multiband surveys, the sample is not radio-selected since it
  includes both RQ and RL AGN (20 and 17 sources, respectively).

\begin{figure*}
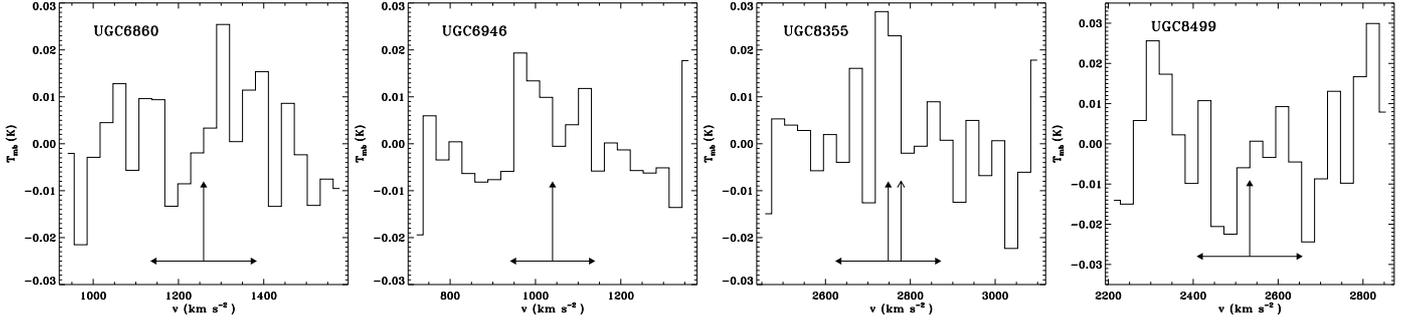

\centerline{
\includegraphics[scale=0.25,angle=90]{6860nroCO1.epsi}
\includegraphics[scale=0.25,angle=90]{6946nroCO1.epsi}
\includegraphics[scale=0.25,angle=90]{8355nroCO1.epsi}
\includegraphics[scale=0.25,angle=90]{8499nroCO1.epsi}
}
\caption{CO(1-0) spectra of the objects observed with the NRO telescope. In
  each panel, the vertical arrow with filled head represents the central
  position of the line used to perform the Gaussian fit, while the other arrow
  marks the position expected by galaxy's recession velocity. The horizontal
  arrows represent the velocity range on which the spectrum is integrated to
  derive the flux measurement.}
\label{spectraCO1nro}
\end{figure*}

\begin{table*}
\begin{center}
  \caption{Results of the CO observations on our sample of ETGs}
\begin{tabular}{c c | c | c c c c | c c c | c | c}
  \hline
  &                 &           & \multicolumn{4}{c}{$^{12}$CO(1-0)} & \multicolumn{3}{c}{$^{12}$CO(2-1)}   &             &\\
  Galaxy Name&     V          & Telescope & $F_{CO}$     & $\Delta v$ & rms & Profile & $F_{CO}$ & $\Delta v$ & rms   & M$_{H_{2}}$ & CO ratio\\
  & km s$^{-1}$ &           & K km s$^{-1}$&km s$^{-1}$ & mK  &         & K km s$^{-1}$ & s$^{-1}$ & mK & 10$^8$ M$_{\odot}$ &\\ 
  \hline
  UGC0968 & 2379       & IRAM      &  3.45$\pm$0.58 &   430    & 7.2 &  dh     &  5.9$\pm$1.2  &  430  & 10  &  1.48$\pm$0.25 & 0.60$\pm$0.2\\
  UGC5617 & 1151       & IRAM      &  1.40$\pm$0.40&   304    & 5.9 &  sp     &  3.03$\pm$0.37 &  300  & 3.8 &  0.31$\pm$0.09 & 0.75$\pm$0.2\\
  UGC6860 & 1259       & NRO       &  $<$9.4       &   -    & 11.0&         &                &       &     &  $<$0.85 & \\
  UGC6946 & 1040       & IRAM      &  $<$4.8       &   -    & 4.1 &         & $<$1.2        &  -  & 3.7 &  $<$0.71 &  \\
  UGC6946 & 1040       &  NRO      & 2.5$\pm$0.7  &   187    & 9.0 &  dh      &                &       &     &  0.17$\pm$0.05 &\\
  UGC7203 & 2215       & IRAM      &  $<$4.5       &   -    & 4.5 &         & $<$1.4        &  -  & 4.6 &  $<$2.95 &\\
  UGC8355 & 2778       & NRO       &   2.3$\pm$0.7 &   240    & 8.2 &  sp     &                &       &     &  0.91$\pm$0.28 &\\
  UGC8499 & 2533       & NRO       &  $<$10.5     &   -     & 14.9&         &                &       &     &  $<$3.6 &\\
  UGC9706 & 1714       & IRAM      &  $<$4.3      &   -     & 3.7 &         &  $<$2.9       &  -  & 7.3 &  $<$1.6 &\\
  \hline
\end{tabular}
\label{tab1}
\end{center}
\medskip

Column description: (1) name; (2) recession velocity; (3) telescope used for
the observation; CO(1-0) observations: (4) integrated line emission in K km
s$^{-1}$; (5) velocity width; (6) rms of the spectrum in mK; (7) shape of the
line profile: dh (double-horn), sp (single peak); CO(2-1) observations: (8)
integrated line emission in K km s$^{-1}$; (9) spectral region of integration; (10) rms of
the spectrum in mK; (11) evaluation of H$_{2}$ mass from CO(1-0) line
intensity; (12) CO(2-1)/CO(1-0) ratio.
\end{table*}

\begin{figure}
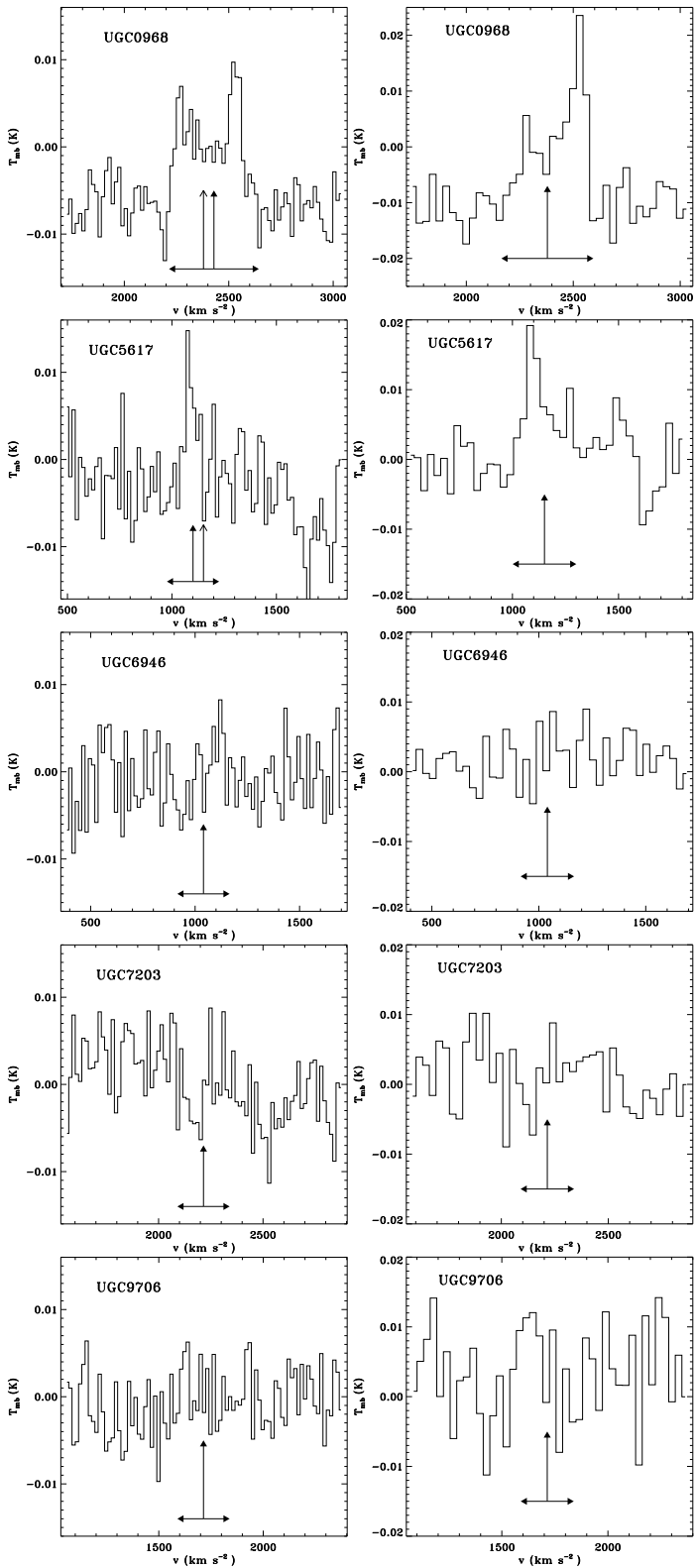

\centerline{
\includegraphics[scale=0.25,angle=90]{968iramCO1.epsi}
\includegraphics[scale=0.25,angle=90]{968iramCO2.epsi}
}
\centerline{
\includegraphics[scale=0.25,angle=90]{5617iramCO1.epsi}
\includegraphics[scale=0.25,angle=90]{5617iramCO2.epsi}
}
\centerline{
\includegraphics[scale=0.25,angle=90]{6946iramCO1.epsi}
\includegraphics[scale=0.25,angle=90]{6946iramCO2.epsi}
}
\centerline{
\includegraphics[scale=0.25,angle=90]{7203iramCO1.epsi}
\includegraphics[scale=0.25,angle=90]{7203iramCO2.epsi}
}
\centerline{
\includegraphics[scale=0.25,angle=90]{9706iramCO1.epsi}
\includegraphics[scale=0.25,angle=90]{9706iramCO2.epsi}
}
\caption{CO(1-0) (left) and CO(2-1) (right) spectra of the objects observed
  with the IRAM 30m telescope. In each panel, the vertical arrow with filled head represents the
  central position of the line used to perform the Gaussian fit, while the
  other arrow marks the position expected by galaxy's recession velocity. The
  horizontal arrows represent the velocity range on which the spectrum is
  integrated to derive the flux measurement.}
\label{spectraCO1iram}
\end{figure}

We proposed to observe with the Nobeyama Radio Observatory (NRO) 45 m
telescope in Nobeyama (Japan) and with IRAM 30 m telescope in Pico
Veleta (Spain) the sources of the sample which do not have previous CO
detections. In this work we report the CO data for the eight sources
whose we got observing time in April and October-November 2007,
respectively for the NRO and IRAM telescopes.  Four galaxies were
observed with the NRO telescope covering the CO(1-0) line at 115 GHz,
and five with the IRAM 30m telescope to detect the CO(1-0) line as
well as the CO(2-1) line at 230 GHz. UGC~6946 was observed with both
telescopes. The beam full width at half maximum (FWHM) of NRO at 115
GHz is 15\arcsec, while those of the IRAM 30m telescope are,
respectively, 22\arcsec\, and 13\arcsec\, at the two CO lines
frequencies. The observations at NRO were carried out with S100
receiver and with Auto Correlator (AC) spectrometer with a band width
of 512 MHz, whereas those at IRAM 30m telescope using the wobbler
switching mode with band widths of 512 and 1024 MHz, respectively, for
the two frequencies.

For NRO we spent $\sim$5 hours for single object to reach an RMS noise
of $\sim$3 mK for $\sim$30 km s$^{-1}$ channel. Instead the total
integration time for each source observed with the IRAM 30m telescope
was $\sim$1--3 hours, resulting in a relatively homogeneous level of
noise of 2 mK per 15 and 30 km s$^{-1}$ channel, respectively, for the
two CO frequencies. The signal is expressed in main beam temperature
T$_{\rm mb}$, since the sources are not expected to be extended and
homogeneous. T$_{\rm mb}$ is related to the equivalent antenna
temperature reported above the atmosphere (T$^{*}_{A}$) by
$\eta=$T$^{*}_{\rm A}/$T$_{\rm mb}$, where $\eta$ is the telescope
main-beam efficiency.  For IRAM telescope, at 115\,GHz $\eta$ = 0.79
and at 230 GHz $\eta$ = 0.54, while for NRO telescope $\eta=$0.34 at
115\,GHz. Each spectrum was summed and reduced using linear baselines.

\section{Results}
 \subsection{CO spectra}

Let us now present the new CO observations of the 8 objects:
Fig.~\ref{spectraCO1nro} shows the NRO spectra for the CO(1-0)
rotational transition for four objects and Fig.~\ref{spectraCO1iram}
shows the IRAM spectra for both the CO(1-0) and CO (2-1) lines for
five objects. We detect the CO(1-0) and CO(2-1) lines for 4 (out of 9)
and 2 (out of 5) of the observed objects, respectively.

Since some lines have complex profiles, to yield a correct measurement of the
flux for the 3-$\sigma$ detected line, we directly integrate the T$_{mb}$
profile on the spectrum for all the sources:
$$
I_{CO} = {\textstyle \int T_{mb}(v)\,dv}    \,\,\,\,\,\,\,\, [\mbox{K km s}^{-1}]
$$ For the undetected CO lines, assuming a typical 250 km s$^{-1}$
line width (e.g., \citealt{young11}), corresponding to a good sampling
of the line with at least 8 (16) points in the spectrum with 30 (15)
km s$^{-1}$ of spectral resolution for NRO (IRAM) telescope, the
1$\sigma$ upper limits are evaluated in this way:
$$
I_{CO} = \sqrt{N_{ch}} \,\, RMS_{T_{mb}} \,\, \Delta v 
$$
where N$_{ch}$ is the number of channel covered in 250 km s$^{-1}$ line width
($\Delta v$) and RMS$_{T_{mb}}$ is the temperature rms of the spectrum.  

Among the 4 sources showing CO(1-0) detection, UGC~0968
and UGC~6946 exhibit double-horn profiles, while UGC~5617 and UGC~8355 have
single-peak profiles. Galaxies detected in the CO(2-1) line (UGC~0968 and
UGC~5617) have profiles similar to those found in the CO(1-0) line. In
Tab.~\ref{tab1}, we collect the CO(1-0) and CO(2-1) line properties for the
observed objects.

Among the objects of the sample, a few have additional CO measurement in the
literature. For UGC~5617, \citet{lisenfeld08} measured both the CO lines with
IRAM telescope: I$_{CO(1-0)}$ = 1.3 K km s$^{-1}$, consistent with our IRAM CO
flux and I$_{CO(2-1)}$ = 0.7 K km s$^{-1}$, slightly smaller than our
measurement. Our non-detections of CO(1-0) for UGC~8499 and UGC~9706 are also
consistent with the upper limits reported by \citet{combes07} with IRAM
telescope, $<$2.37 and $<$4.8 K km s$^{-1}$, respectively.

\subsection{CO emission and H$_{2}$ mass estimation}
\label{COH2conversion}

From standard, empirical calibrations, it is possible to deduce the
interstellar H$_{2}$ mass content from the integrated CO intensity
I$_{CO}$ (K km s$^{-1}$). We adopt here the conversion factor
commonly used for $$N(H_{2}) = 2.3 \times 10^{20} \,\, I_{CO} $$ in
molecule/cm$^{2}$ unit. From this equation, the mass of cold molecular
hydrogen contained in one beam of the telescope is:
$$
M_{H_{2}} = \alpha \,\, I_{CO} \,\, A_{beam}  
$$ 
where $\alpha$ is the conversion factor, whose commonly used value is 4.8
M$_{\odot}$ (K km s$^{-1}$ pc$^{2}$)$^{-1}$ \citep{solomon91}, derived from
galactic molecular cloud observations, and A$_{beam}$ is the beam area at the
galaxy in pc$^{2}$.

The conversion between CO and M$_{H_{2}}$ column densities/masses
remains controversial. Its value varies with the physical gas
conditions, metallicity, and environmental factors (e.g.,
\citealt{maloney90,israel97,boselli02,bolatto08,glover11,magrini11})
and can change by a factor 4-15 (e.g.,
\citealt{wilson95,casasola07}). The conversion H$_{2}$-CO value also
changes with galaxy morphological type: galaxies earlier than Scd type
usually show values comparable to, or lower than, the Galactic one,
while extremely late-type spirals or irregular galaxies tend to show
higher values \citep{arimoto96}.

The H$_{2}$ can be underestimated if the molecular cloud is more
extended than the beam area covered on the galaxy. In our case such a
physical region at the distance of the sources corresponds to an
aperture which ranges from 0.5 to 4 kpc for the CO(1-0) transition. CO
interferometric observations show that the molecular gas in ETGs is
centrally concentrated in the central kiloparsecs (e.g.,
\citealt{braine93,boselli02,boker03,komugi08,davis13}). Recently,
\citet{alatalo13}, by comparing the CO fluxes recovered from 30m
spectra and from full interferometric maps (from CARMA) for a sample
of local ETGs, found that up to half of the molecular mass of a source
(at a similar distance to our sample) measured with the 30m telescope
can be missed.  However, no large differences are expected between the
30m and 45m telescope observations. Therefore, this effect should not
effect significantly our results.

With the method explained above, the H$_{2}$ masses estimated for the eight
galaxies are $7.2 \lesssim {\rm log} \, M_{\rm{H_2}}/M_{\odot} \lesssim 8.5$,
and all values, including the CO upper limits, are listed in
Tab.~\ref{tab2}.

\subsection{CO(2-1)/CO(1-0) ratio}

The CO(2-1)/CO(1-0) is measured by the ratio between the integrated
emission of the two lines on equal regions of the galaxy. Because of
the different beam sizes at 115 and 230 GHz, a beam dilution factor is
necessary to correct the different spatial resolution, that
corresponds to $\sim$4, the ratio of the beam areas for the two lines,
assuming the gas distribution of both CO(3-2) and CO(1-0) is
compact. Therefore the CO(1-0) lines intensities have been multiplied
by this beam correcting factor to be compared to the CO(2-1) lines. It
is possible to evaluate this ratio only for two objects, namely
UGC~0968 and UGC~5617 (Tab.~\ref{tab1}). For UGC~0968 the assumption
of a compact CO distribution is less obvious, since one peak of the
CO(2-1) horn profile is less pronounced than in CO(1-0). This might
suggest that the H$_{2}$ emitting region is intermediate between the
two beam sizes.  However, both values are consistent within the
uncertainties with the CO(2-1)/CO(1-0) line ratio ($\sim$0.6) found
for samples of radio galaxies \citep{ocana10} and Seyfert galaxies
\citep{papadopoulos98}.

\begin{table*}
  \caption{Multiwavelength properties of the whole sample.}
\begin{center}
\begin{tabular}{c| c | c c|c c c c | c | c}
\hline
Name &  z   &  $M_{\rm H_{2}}$ &  Ref. &$L_{\rm r}$ & $L_{\rm o}$&  $L_{\rm X}$ & $L_{[O~III]}$  & class & dust\\
\hline	                              
UGC~0968 & 0.007935  & 8.17   & this work & 36.96 &$<$39.77&  $-$    & $<$37.75          & CoreG & Y \\
UGC~5902 & 0.003039  & $<$6.70& CO07 & 35.83 &$<$39.10&$<$38.40 &    37.96       &    CoreG & N \\
UGC~6297 & 0.003202  & 8.37   & WE03 & 36.84 &$<$39.06&$<$38.40 &    38.21                     & CoreG & Y \\
UGC~7203 & 0.007388  & $<$8.47& this work & 37.44 & 39.85  &$<$38.93 &    38.08           & CoreG & N \\
UGC~7360 & 0.007465  & $<$7.84& OC10 & 39.22 & 39.71  & 40.95   &    38.91                 & CoreG & N \\
UGC~7386 & 0.002165  & 7.36   & CO07 & 38.38 & 38.76  & 39.72   &    38.95                       & CoreG & N \\
UGC~7494 & 0.003536  & $<$7.51& WI95 & 38.57 & 39.27  & 39.30   &    38.20                   & CoreG & N \\
UGC~7629 & 0.003326  & 7.60   & HU94 & 36.73 & 38.79  & 38.23   &    37.40                      & CoreG & Y \\
UGC~7654 & 0.004360  & 7.72   & OC10 & 39.90 & 40.72  & 40.30   &    38.99                      & CoreG & N \\
UGC~7760 & 0.001134  & $<$7.23& CO07 & 37.30 & 38.73  & 38.40   &    37.28                   & CoreG & Y \\
UGC~7797 & 0.006605  & 7.96   & SO93 & 38.05 &$<$40.19& $-$     &    38.79                     & CoreG & Y \\
UGC~7878 & 0.003129  & $<$6.85& SA07 & 36.90 & 39.07  & $<$38.41&    37.89               & CoreG & Y \\   
UGC~7898 & 0.003726  & 7.80   & SA89 & 37.46 &$<$39.13& $<$38.52& $<$37.34              & CoreG & N \\
UGC~8745 & 0.005851  & $-$    &      & 37.81 &  $-$   & $-$     &    38.41                                & CoreG & Y \\
UGC~9655 & 0.006578  & $<$7.46& CO07 & 36.96 &  $-$   & $-$     & $<$38.04                   & CoreG & Y \\
UGC~9706 & 0.005717  & $<$8.20& this work & 37.31 & 39.25  & 38.26   &    38.18                    & CoreG & Y \\ 
UGC~9723 & 0.002242  & 7.92   & TA94 & 36.92 & $-$    & $<$38.18&    37.33                            & CoreG  & Y \\            
\hline 		    		    
UGC~5617 & 0.003839  & 7.49   & this work & 36.88  & $<$40.51  & 40.30   & 38.75        & PlawG & Y  \\ 
UGC~5663 & 0.004383  & $<$8.31& SA89 & 36.87  &	40.57  & 38.80   & 38.59               & PlawG  & Y \\ 
UGC~5959 & 0.004717  & $<$6.95& CO07 & 37.16  &	40.54  & 39.80   & 38.77                & PlawG  & Y \\ 
UGC~6153 & 0.008836  & 8.61   & MA97 & 38.14  &	42.11  & 42.40   & 40.80               & PlawG & Y  \\ 
UGC~6860 & 0.004200  & $<$7.93& this work & 36.38  & $<$40.15  & 39.60   & 38.28      &  PlawG & N  \\ 
UGC~6946 & 0.003469  & 7.24   & this work & 38.16  &	41.40  & 41.00   & 39.42       & PlawG & Y \\ 
UGC~6985 & 0.003102  & 7.94   & WE03 & 36.31  & $<$40.56  & $-$     & $<$37.51         & PlawG & N \\ 
UGC~7005 & 0.004820  & $-$    &      & 36.94  & $<$40.37  & 40.40   & 39.09                      & PlawG  & Y \\ 
UGC~7103 & 0.002692  & 7.20   & WE03 & 36.39  &	40.85  & 39.40   & 38.45                       & PlawG  & N \\ 
UGC~7142 & 0.003196  & $<$6.65& WE03 & 36.96  &	40.44  & 39.90   & 38.71                       & PlawG  & Y \\ 
UGC~7256 & 0.003623  & 6.88   & WE03 & 37.33  &	40.57  & 40.20   & 39.01             & PlawG  & Y \\ 
UGC~7311 & 0.007909  & $-$    &      & 37.07  &	$-$    & $-$     & 38.35                           & PlawG  & Y \\ 
UGC~7575 & 0.002672  & 7.98   & VO05 & 36.00  &	$-$    & 39.30   & 38.72               & PlawG & Y   \\ 
UGC~7614 & 0.004036  & 8.23   & CO07 & 36.15  &	40.47  & $<$40.40& $<$37.85    & PlawG & Y \\  
UGC~8355 & 0.009267  & 7.96   & this work & 36.69  &	41.27  & $-$  & $<$38.31       & PlawG & N \\ 
UGC~8499 & 0.008449  & $<$8.56& this work & 37.09  & $<$40.37  & $-$     & 38.31      &  PlawG  & N \\ 
UGC~8675 & 0.003549  & $<$7.90& MA97 & 36.34  &	40.33  & 40.60   & 39.66             & PlawG & Y \\ 
UGC~9692 & 0.004533  & $<$7.34& CO07 & 36.65  & $<$40.41  & $-$     & 37.92         & PlawG & Y \\
UGC~10656& 0.009306 & $-$    &      & 37.05  & $<$40.96  & 40.10   & 38.90                    & PlawG & N \\
UGC~12759& 0.005704 & 8.12   & MA97 & 36.93  &	40.49  & 41.47   & 39.51                & PlawG &  N \\
\hline
\end{tabular}

\medskip
\end{center}

Column description: (1) name; (2) redshift; (3) $H_{2}$ mass in M$_{\odot}$ and references
(4): CO07 \citet{combes07}, WE03 \citet{welch03}, OC10 \citet{ocana10}, WI95
\citet{wiklind95}, HU94 \citet{huchtmeier94}, SO93 \citet{sofue93}, SA07
\citet{sage07}, SA89 \citet{sage89}, TA94 \citet{taniguchi94}, MA97
\citet{maiolino97}, VO05 \citet{vollmer05}; nuclear luminosities [erg
s$^{-1}$] from \citet{balmaverde06b} and \citet{capetti06} in (5) radio, (6)
optical, and (7) and X-ray (2-10 keV); (8) [O~III] emission line luminosity [erg
s$^{-1}$] from  \citet{capetti06}, \citet{baldi09}; (9) optical profile; (10) presence of dust structure in HST optical image? Y for
yes and N for no.
\label{tab2}
\end{table*}

\begin{figure}
\centering{
\includegraphics[scale=0.43]{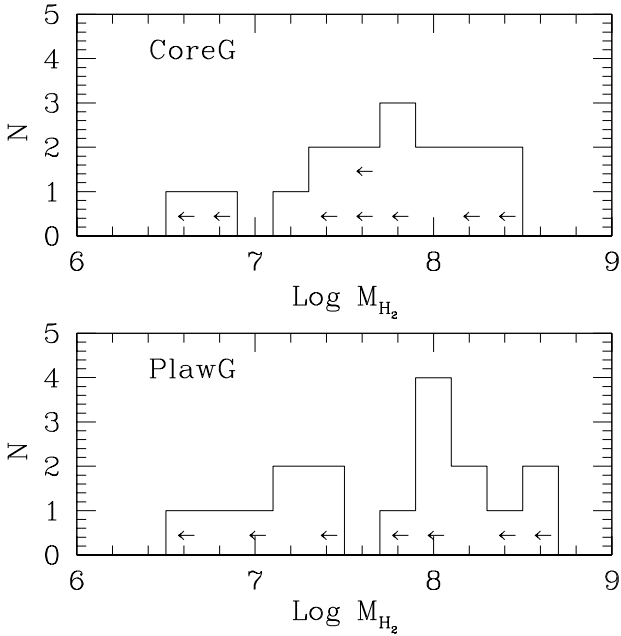}
}
\caption{Histograms of the $M_{H_{2}}$ in M$_{\odot}$ units for the
  CoreG (top panel) and PlawG (bottom panel). The arrows indicate the
  upper limits on the molecular mass measurement.}
\label{isto}
\end{figure}

\section{H${_2}$ mass for the sample of nearby ETGs}

We collect from the literature the $H_{2}$ masses, estimated from
CO(1-0) observations, of the remaining objects of the sample studied
by \citet{capetti06}. Combining our new results and the literature
data, we now have CO observations for all but 4 (out of 37)
sources. Tab.~\ref{tab2} shows the values of molecular masses and the
appropriate references for the entire samples of CoreG and PlawG.

The overall CO detection rate is 55$\pm$9\% and it is similar
for the two classes of brightness profiles (50$\pm$12\% for the
CoreG and 59$\pm$12\% for the PlawG, respectively). The
distributions of the molecular masses for the two sub-samples (left
panel in Fig.~\ref{isto}) is 10$^{6.5}$ - 10$^{8.5}$ M$_{\odot}$. This
broad range is consistent with the molecular masses found in other
samples of ETGs, on average, of several 10$^{7}$ M$_{\odot}$ (e.g.,
\citealt{lees91,wiklind95,combes07,crocker11,young11}).

Given the presence of no-detections in the data, we use the
Kaplan-Meier product-limit estimator \citep{kaplan58} which provides
the mean values for distribution of $M_{H_{2}}$ with the assumption
that the censoring is random, condition fulfilled in our data. For
CoreG, $\langle \log M_{H_{2}} \rangle$ = (7.4 $\pm$ 0.2) M$_{\odot}$ and for PlawG,
$\langle \log M_{H_{2}} \rangle$ = (7.5 $\pm$ 0.2) M$_{\odot}$. The two average
values are consistent with each other within the errors.

To asses the presence of significantly statistical difference between
the $M_{H_{2}}$ distributions of CoreG and PlawG, we use the
`twosampt' task of the Astronomy Survival Analysis (ASURV) package
\citep{lavalley92}. This package is available under IRAF/STSDAS and
provides us a tool to deal with censored data (\citealt{feigelson85},
\citealt{isobe86}). In particular, the Peto-Prentice test
\citep{peto72} quantifies the distance between the empirical
distribution functions of two samples, with the null hypothesis that
they are drawn from the same distribution. The result of the test is
that the M$_{H_{2}}$ distribution of CoreG and PlawG are not drawn from
different parent populations with a confidence level greater than
95\%.

\begin{figure}
\centering{
\includegraphics[scale=0.43]{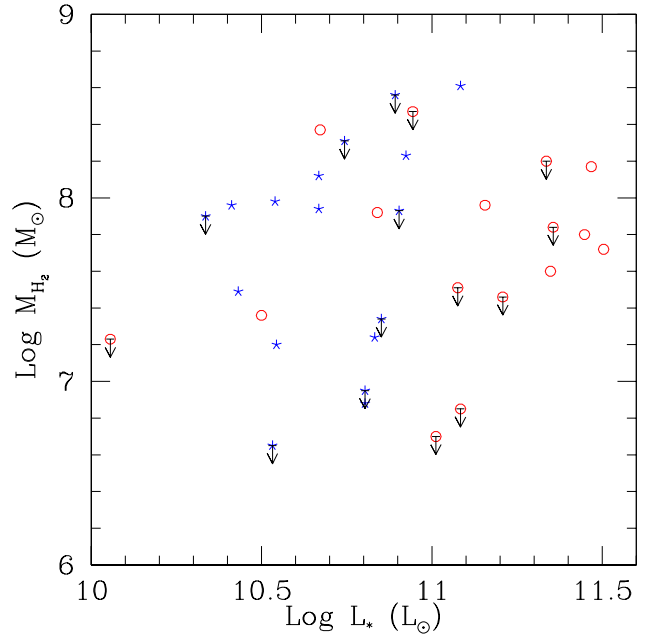}
}
\caption{Luminosity of the host galaxies estimated from the K-band
  magnitude (in L$_{\odot}$ units) vs.\ $H_{2}$ masses (in M$_{\odot}$
  units). The RL AGN (CoreG) are the red empty circles, while the RQ
  AGN (PlawG) are the blue stars.}
\label{LMplane}
\end{figure}

In Fig.~\ref{LMplane} we compare the molecular masses and
stellar luminosity, $L_{*}$, estimated from the K-band absolute magnitude of
the galaxy. The contribution of molecular mass is in the range $\sim 10^{-2} -
10^{-4}$ of the galaxy luminosity (corresponding to the galaxy
mass). \citet{capetti06} found that the CoreG hosts are more luminous and so
more massive than those of the PlawG. Yet, the Peto-Prentice test keeps on
confirming that the CoreG and PlawG are not significantly different in terms
of fraction $M_{H_2}/L_{*}$, with a confidence level greater than 95\%.

We further investigate the presence of a difference between CoreG and
PlawG in the HST optical images \citep{capetti05} to look for an
association between the presence of dust and the CO detection. We
exclude the 4 objects (UGC~7005, UGC~7311, UGC~8745, and UGC~10656)
that have no CO observations (Tab.~\ref{tab2}). Most of the objects
(20/33) shows evident optical obscuration in the nuclear region. Among
the 20 objects with dust presence in their hosts, 11 have a CO
detection. Among the 13 objects without dust structure, 7 have a CO
detection. This implies that there is no a clear association between
the presence of dust and the CO detection in the ETGs of our
sample. In addition, the dust is indifferently present in CoreG and
PlawG.

\subsection{Molecular gas and nuclear properties}

\begin{figure*}
\centering{
\includegraphics[scale=0.43]{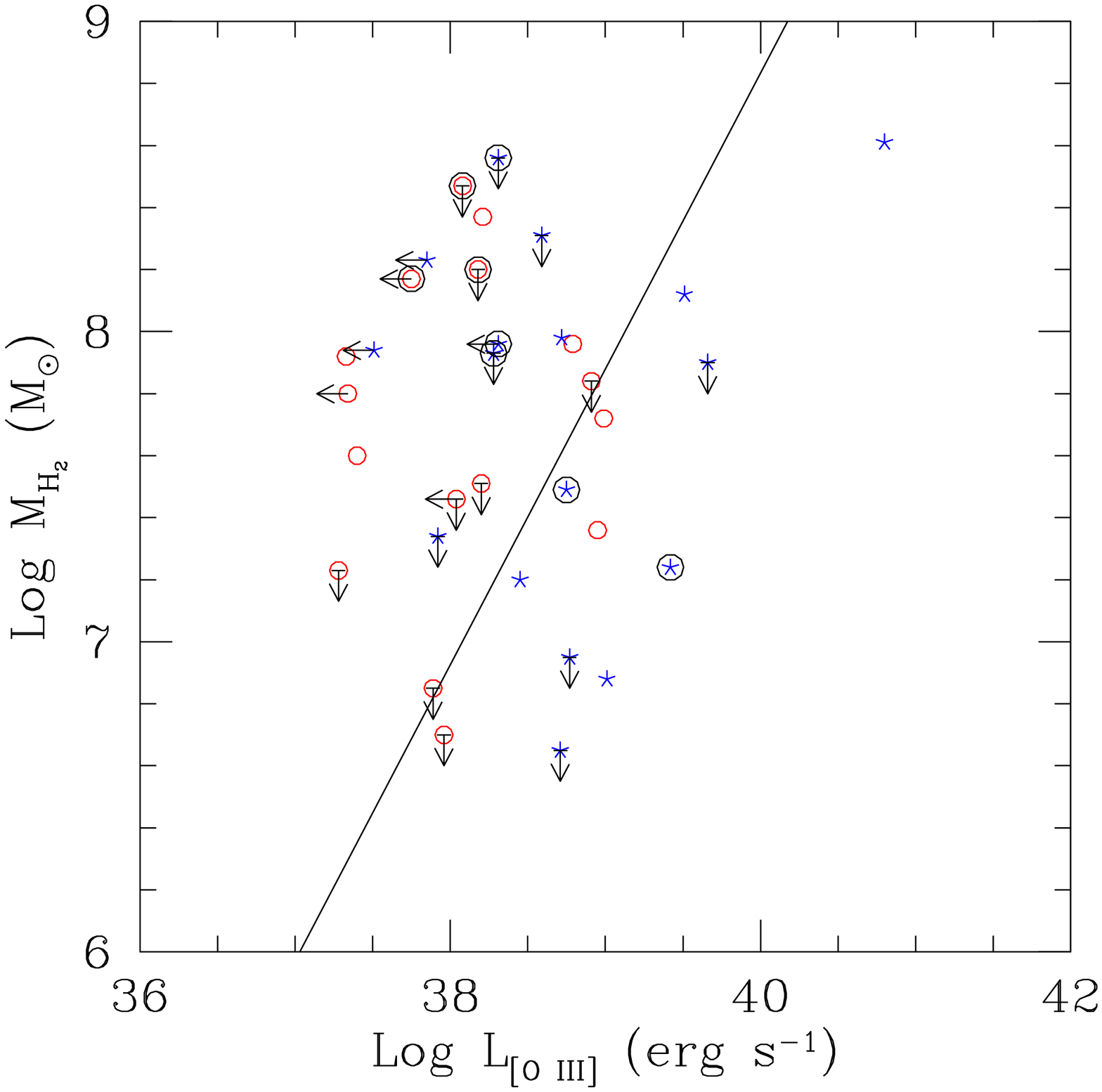}
\includegraphics[scale=0.43]{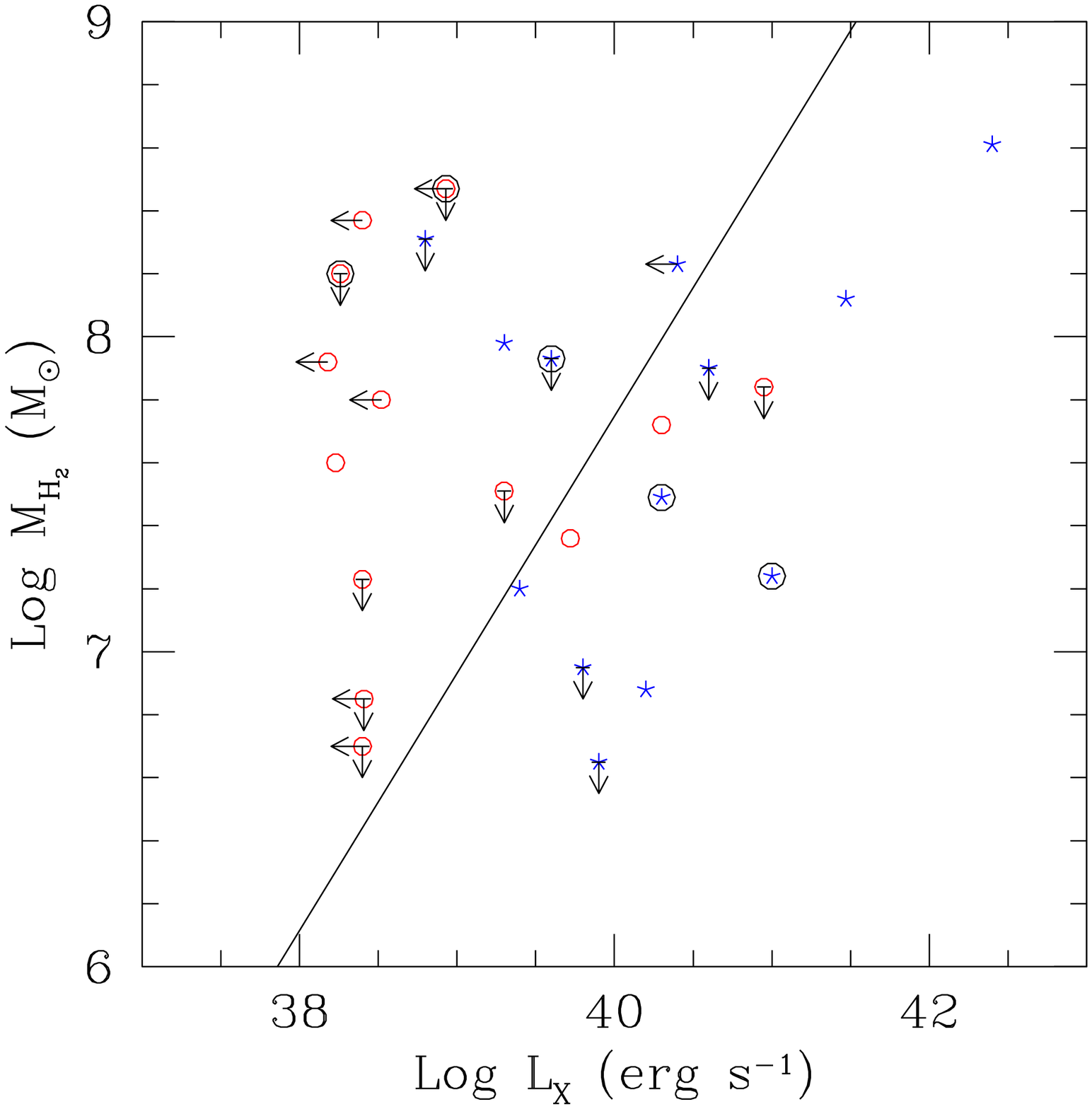}
}
\centering{
\includegraphics[scale=0.43]{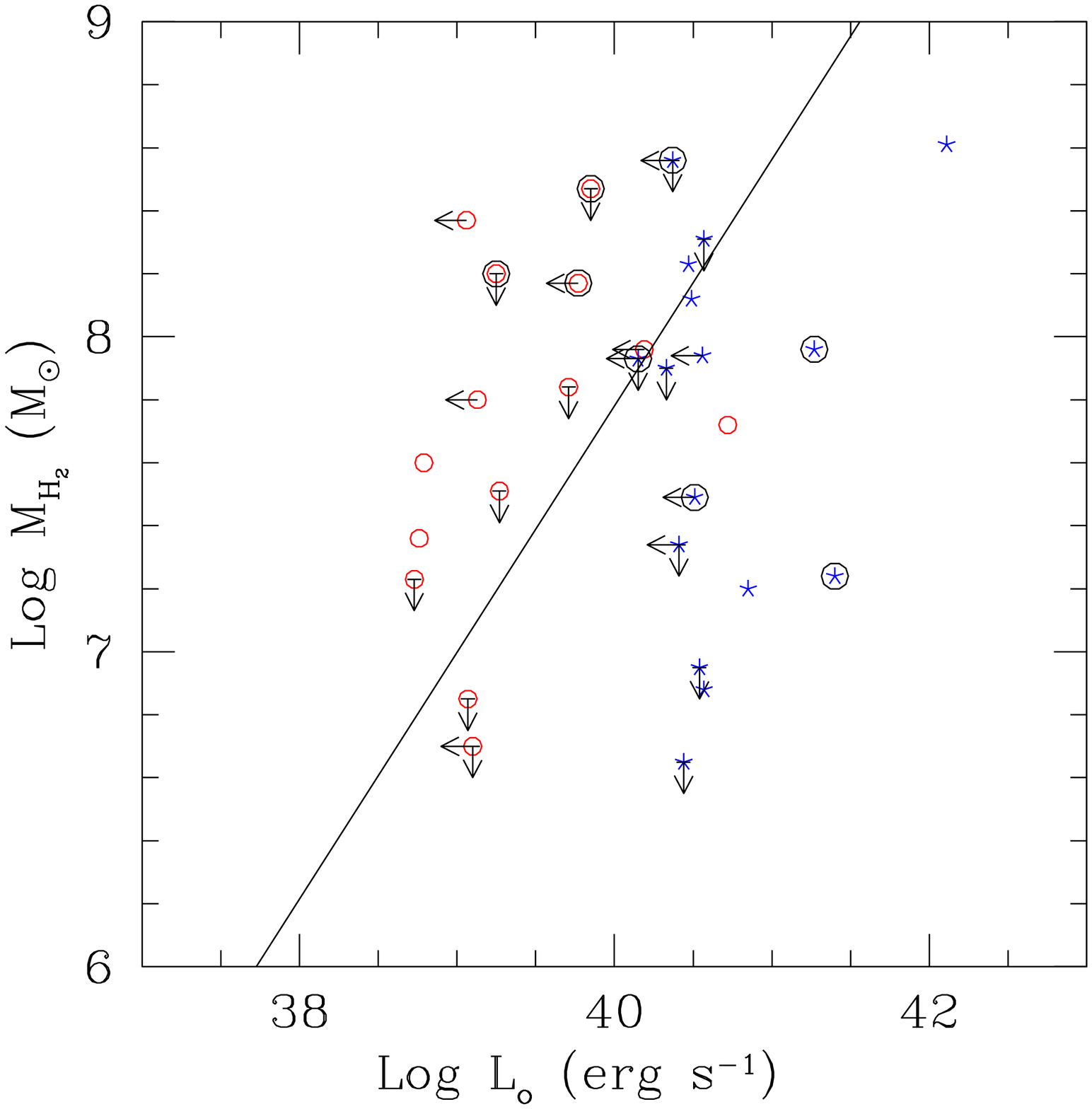}
\includegraphics[scale=0.43]{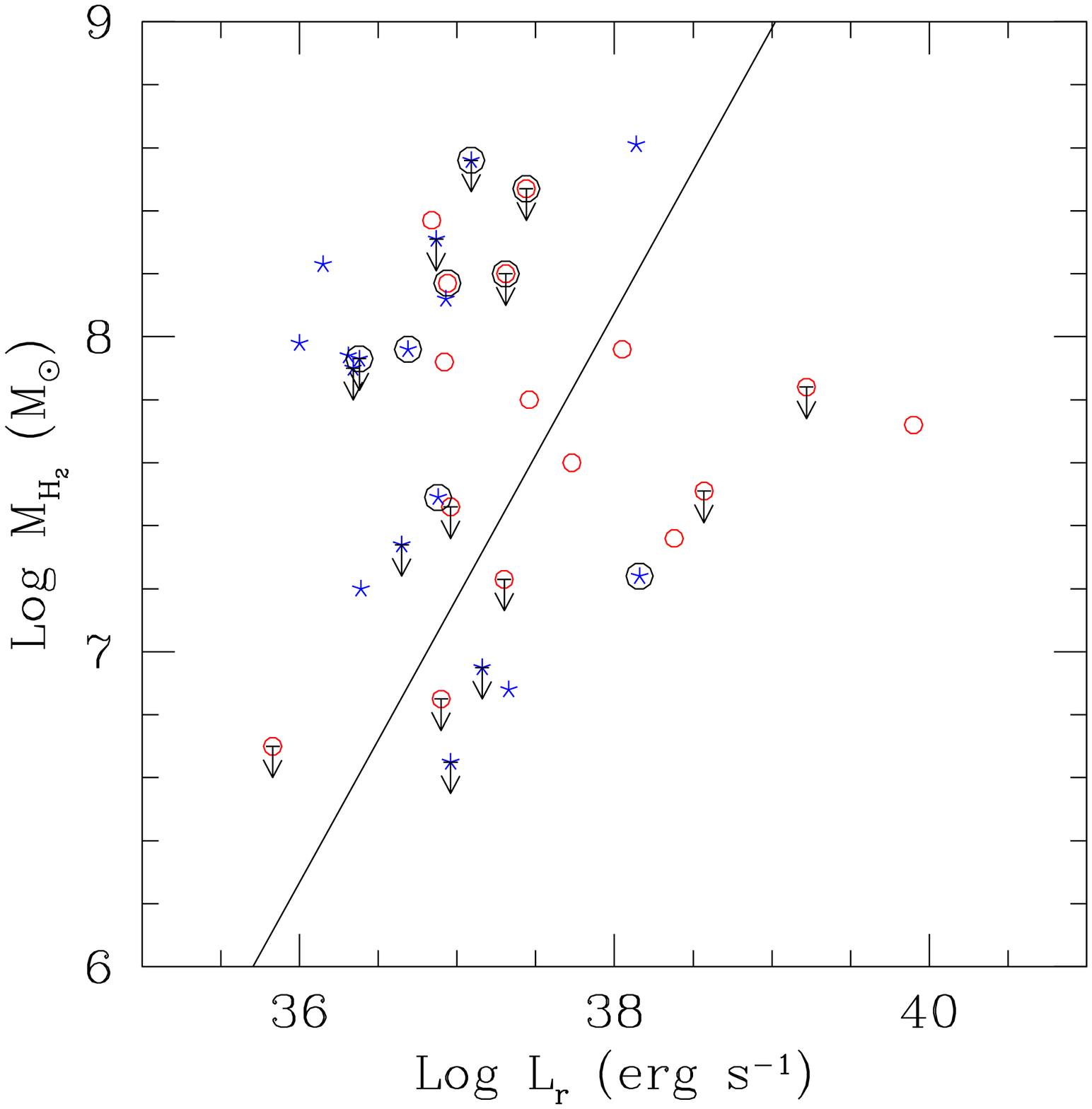}
}
\caption{Multiwavelength luminosities (in erg s$^{-1}$) versus $H_{2}$
  masses (in M$_{\odot}$). Upper left panel: [O~III] luminosities
  vs. $M_{H_{2}}$; upper right panel: nuclear X-ray luminosities
  vs. $M_{H_{2}}$; lower left panel: optical luminosities
  vs. $M_{H_{2}}$; lower right panel radio core luminosities (5 GHz)
  vs. $M_{H_{2}}$. The RL AGN (CoreG) are the red empty circles, while
  the RQ AGN (PlawG) are the blue stars. The large black circles mark the
  molecular masses estimated from our CO observations.}
\label{plots}
\end{figure*}

Another approach to study the molecular gas content in our sample of
ETGs is to look for a possible connection with their nuclear
properties by considering the multiwavelength emission from their
AGN. Operatively, we compare M$_{H_{2}}$ with various tracers of
nuclear activity, in terms of radio (at 5 GHz), optical, X-ray (2-10 keV) and
[O~III] emission line nuclear power for the entire sample (see panels
in Fig.~\ref{plots}).

At a first glance, the panels of Fig.~\ref{plots} shows 'trends', with
relatively large scatters, between the H$_{2}$ mass and different
indicators of the AGN power. Our new CO measurements are, on average,
located at large molecular masses and intermediate nuclear
luminosities. However, it is interesting to point out that UGC~6946
which has the lowest $H_{2}$ mass among the new CO detection is one of
the most luminous sources in [O~III], X-ray, optical, and radio of the
sample.

To assess the presence of correlations, we performed a statistical
censored analysis using the ASURV package. Note that the method we use
is based on the assumption that the censored data are uniformly
distributed, condition that is satisfied for our data. We used the
``schmittbin'' task \citep{schmitt85} to calculate the associated
linear regression coefficients for two sets of variables. Operatively,
we carried out this procedure twice, obtaining two linear regressions:
first, we considered the former quantity as the independent variable
and the latter as the dependent one; second, we switched the roles of
the variables. The best fit is represented by the bisector of these
two regression lines. This follows the suggestion of \citet{isobe90}
that consider such a method preferable for problems that require a
symmetrical treatment of the two variables. In order to estimate the
quality of the linear regression, we derived the generalized Kendall's
$\tau$ \citep{kendall83} between the two variables, using the
``bhkmethod'' task. The statistical parameters of all the linear
regressions are reported in Tab.~\ref{statist1}.

Although the low statistical parameters, the presence of a weak trend
between the [O~III] line luminosity and $M_{H_{2}}$ (upper left panel
of Fig.~\ref{plots}) can be marginally considered. The Kendall's
$\tau$ coefficient is 0.140 with an associated probability, that a
fortuitous correlation appears at a level measured by our test
statistic, of 0.42. For the other tracers of the nuclear activity,
X-ray, optical and radio band (right upper panel and lower panels of
Fig.~\ref{plots}), the statistical significance of a relation with the
molecular gas mass is lower. In fact, the probability which a
fortuitous relation appears in those plots are 0.65, 0.69, and 0.67,
respectively (Table~\ref{statist1}). In the $L_{radio}$-$M_{H_{2}}$
panel the lack of the relation is also probably driven by the the
different radio-loudness of CoreG and PlawG. We cannot establish the
type of relation between L$_{AGN}$ and M$_{H_{2}}$ since the slopes of
the trends are not well constrained due to the low statistics.

Although caution is advised, the significant scatter observed in the
observed trends M$_{H_{2}}$-L$_{AGN}$ suggests that, if a connection
between nuclear properties and molecular gas content is present, it is
certainly affected by complex mechanisms. The lack of clear starburst
regions in the optical HST images \citep{capetti05} and no evidence of
young stellar populations in their optical spectra \citep{baldi09}
rule out the star formation origin of these trends.  A reasonable, but
not obvious, explanation of the trends might be the most basic
mechanism of heating of the molecular clouds by high-energy photons
(e.g., UV and X-ray photons) from the AGN, as suggested by
\citet{braine92}. The increase of the AGN luminosity might account for
the increase of the CO luminosity, i.e. the molecular mass heated by
the radiation field and, then, emitting CO lines. The possible sources
of the production of high-energy photons are the accretion disk and
the relativistic radio jet. The contribution of the former is
important for the (RQ) PlawG, while the latter for the (RL) CoreG
\citep{capetti06,balmaverde06b}. This explanation is in line with the
results of \citet{nesvadba10} and \citet{ogle10} for RL AGN,
consistent with a model of cold gas emission powered by the
dissipation of mechanical energy through shocks by radio jets.

\begin{table*}
\begin{center}
\caption{Summary of the statistical analysis of the linear regressions for the sample}
\begin{tabular}{c|c|cc|cc|cc} 
\hline\hline
Relation & Class & N$_{d}$ & N$_{c}$  & $\tau$ & P$_{\tau}$ & m & q \\
(1) & (2) & (3) & (4) & (5) & (6) & (7) & (8) \\
\hline
L$_{[O~III]}$-M$_{H_{2}}$ & all   & 13  & 20 & 0.140  & 0.416 & 0.95$\pm$0.30 & -29$\pm$17  \\
L$_{X}$-M$_{H_{2}}$      & all   & 11 & 15 & 0.092  & 0.647 &  0.82$\pm$0.40 &  -25$\pm$20 \\
L$_{o}$-M$_{H_{2}}$      & all   & 11 & 19 &  0.064  & 0.694 & 0.78$\pm$0.40 & -24$\pm$22  \\
L$_{r}$-M$_{H_{2}}$      & all & 17 & 15 &  -0.079 & 0.674 &  0.90$\pm$0.40 &  -26$\pm$17 \\
\hline
\end{tabular}
\label{statist1}
\end{center}

\medskip
Column description: (1) relation studied; (2) sample considered; (3)
number of sources with both the two quantities detected; (4) number of
sources that show, at least, an upper limit in one of the two
quantities; (5) the generalized Kendall's $\tau$, considering all the
data; (6) P$_{\tau}$ probability associated with $\tau$ that the
correlation is not present; (7)-(8) slope coefficient $m$ and intercept
coefficient $q$ for the linear regression ($y = m*x + q$).
\end{table*}

\section{Summary and conclusions}
\label{summary}

We have analyzed the CO(1-0) and CO(2-1) transition spectra for eight
nearby galaxies observed with NRO and IRAM telescopes. These
  sources belong to a sample of 37 ETGs whose nuclear and host
  properties are well studied: they host low-luminosity RL and RQ AGN
associated with flat (CoreG) or sharp (PlawG) optical profile,
respectively. We convert the CO(1-0) luminosities in H$_{2}$ mass
using standard correction factors. To complete the sample, we found in
literature the molecular masses for the remaining of the sample,
ranging from 10$^{6.5}$ to 10$^{8.5}$ M$_{\odot}$.

Despite they differ in nuclear and host properties, we do not find any
statistical significant difference in terms of molecular gas content
between CoreG and PlawG. This suggests that the amount of molecular
gas does not seem to be related to the radio-loudness of the AGN in an
ETG. Since single-dish observations do not provide a spatial
information of the molecular gas distribution, it is
reasonable to propose that the properties of the multi-phase gas at
the nuclear scale, instead of the galaxy scale, sets the accretion
mode and the radio-loudness of the AGN. In fact, at the center of
low-luminosity radio galaxies (which include CoreG), on some parsecs
scale, hot gas is known to be responsible to regulate the nuclear
activity (e.g., \citealt{allen06,balmaverde08,nemmen14a}).

Furthermore, we look for possible connection between the molecular gas
content and the nuclear activity in [O~III] line, X-ray, optical, and
radio band. We do not find highly significant
  correlations. However, the censored analysis indicates that the
  presence of a trend between the H$_{2}$ mass and the [O~III]
  luminosity cannot be excluded. We interpret this link in terms of a
  possible increase in CO luminosity produced by heating of the
  molecular mass by the AGN radiation field which also illuminates the
  narrow-line region. This scenario still needs to be confirmed on a
larger sample of RQ/RL AGN. This will be addressed to a forthcoming
paper. Furthermore, it is fundamental to increase the number of CO
detection to improve the statistics. The ALMA telescope certainly will
play a decisive role in this field in the next future.

\begin{acknowledgements}

This work has benefited from research funding from the European
Community's sixth Framework Programme under RadioNet R113CT 2003
5058187. RDB was supported at the Technion by a fellowship from the
the Lady Davis Foundation. We thank the referee for the constructive
comments that have helped us to improve the paper.


\end{acknowledgements}

\bibliography{my}

\end{document}